\documentclass[sigconf]{acmart}

\acmConference[ICPC 2022]{The 30th International Conference on Program Comprehension}{May 21–22, 2022}{Pittsburgh, PA, USA}

\usepackage{tcolorbox}
\usepackage{multirow}
\usepackage{subfigure}
\usepackage{balance}
\usepackage{ulem}

\def\BibTeX{{\rm B\kern-.05em{\sc i\kern-.025em b}\kern-.08em
    T\kern-.1667em\lower.7ex\hbox{E}\kern-.125emX}}

\copyrightyear{2022}
\acmYear{2022}
\setcopyright{acmcopyright}\acmConference[ICPC '22]{30th International Conference on Program Comprehension}{May 16--17, 2022}{Virtual Event, USA}
\acmBooktitle{30th International Conference on Program Comprehension (ICPC '22), May 16--17, 2022, Virtual Event, USA}
\acmPrice{15.00}
\acmDOI{10.1145/3524610.3527889}
\acmISBN{978-1-4503-9298-3/22/05}

\begin{document}

\title{CSRS: Code Search with Relevance Matching and Semantic Matching}

\author{Yi Cheng}
\affiliation{%
  \institution{Central South University}
  \city{ChangSha}
  \country{China}}
\email{roycheng@csu.edu.cn}

\author{Li Kuang}
\authornote{Li Kuang is the corresponding author.}
\affiliation{%
  \institution{Central South University}
  \city{ChangSha}
  \country{China}}
\email{kuangli@csu.edu.cn}

\begin{abstract}
Developers often search and reuse existing code snippets in the process of software development. Code search aims to retrieve relevant code snippets from a codebase according to natural language queries entered by the developer. Up to now, researchers have already proposed information retrieval (IR) based methods and deep learning (DL) based methods. The IR-based methods focus on keyword matching, that is to rank codes by relevance between queries and code snippets, while DL-based methods focus on capturing the semantic correlations. However, the existing methods do not consider capturing two matching signals simultaneously. Therefore, in this paper, we propose CSRS, a code search model with relevance matching and semantic matching. CSRS comprises (1) an embedding module containing convolution kernels of different sizes which can extract n-gram embeddings of queries and codes, (2) a relevance matching module that measures lexical matching signals, and (3) a co-attention based semantic matching module to capture the semantic correlation. We train and evaluate CSRS on a dataset with 18.22M and 10k code snippets. The experimental results demonstrate that CSRS achieves an MRR of 0.614, which outperforms two state-of-the-art models DeepCS and CARLCS-CNN by 33.77\% and 18.53\% respectively. In addition, we also conducted several experiments to prove the effectiveness of each component of CSRS.
\end{abstract}




\maketitle

\section{Introduction}
In modern society, software systems have already been applied in various fields. Similar functions may exist in different software systems, and their implementation codes are similar to each other. When developing or maintaining software systems, in order to improve the development efficiency, developers tend to search or reuse for existing code in search engines or open source repositories such as Google and GitHub, rather than wasting time redeveloping the function. Therefore, code search has become one of the most frequent activities in software development, and it is necessary to develop a code search tool \cite{b25}.

Many code search methods have been proposed in the past few years to advance the code search task. The existing researches can be divided into two categories: information retrieval (IR) based and deep learning (DL) based \cite{b26}. Earlier code search approaches are mostly based on IR techniques, which focus on keyword matching to calculate the relevant score between a query and a code snippet. For example, Lv et al. \cite{b1} proposed CodeHow, which expands the query with the APIs and performs code retrieval by applying the extended boolean model. Lu et al. \cite{b2} propose an approach that extends a query with synonyms generated from WordNet \cite{b34}.

Obviously, the IR-based methods cannot capture the semantic relationship between the query and the code. To tackle this issue, many DL-based methods have been proposed, which can bridge the semantic gaps between programming language in code and natural language in query. Gu et al. \cite{b3} proposed DeepCS (Deep Code Search), which uses two different LSTM (Long Short-Term Memory) \cite{b6} to jointly embed code snippets and natural language queries into a high-dimensional vector space, and then calculates the cosine similarity between code snippets and queries. Shuai et al. \cite{b4} proposed CARLCS-CNN that leverages convolutional neural network (CNN) \cite{b7} and co-attention mechanism to learn the semantic relationship between code and query.

However, the above two types of methods have different concerns. The characteristic of the IR-based model is that the relevance between queries and code snippets can be determined by the keyword matching signals, and the DL-based model performs well in obtaining the semantic relationship between the query and code. Due to the early IR-based methods being mostly rule-based or heuristic, it can not be well combined with the DL-based model. However, with the development of deep learning, some studies use neural networks to build IR models and apply it to document retrieval \cite{b20,b21,b22,b23,b24}, while this kind of IR method is called neural information retrieval (i.e. neural IR). Therefore, it is possible to obtain the relevance and semantic correlations between queries and code snippets.

\begin{figure}[tbp]
    \centering
	\includegraphics[width=0.48\textwidth]{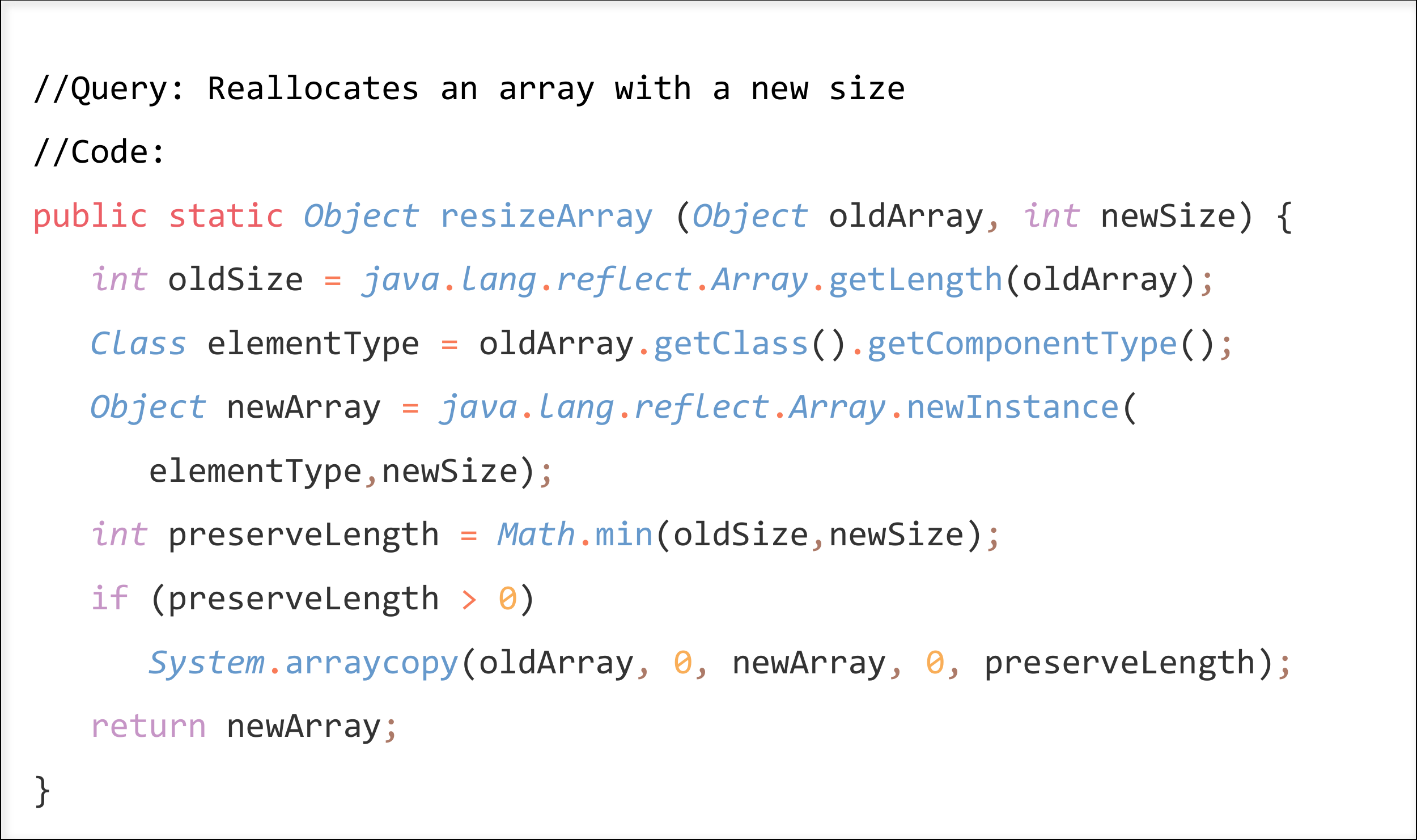}
	\caption{An example of code search.}
	\label{fig:cseg}
\end{figure}

As shown in Fig.\ref{fig:cseg}, the common keywords "array", "new", and "size" are shared in the query and the code, which indicates that the code snippet has relevance to the query. In addition, the query in Fig.\ref{fig:cseg} is semantically correlated with the code. Therefore, in this work, we propose a code search model CSRS (\uline{C}ode \uline{S}earch with \uline{R}elevance \uline{M}atching and \uline{S}emantic \uline{M}atching) to learn the relevance and semantic relationship between query and code. Specifically, CSRS contains a CNN-based encoding module to extract n-gram embedding from query and code, a neural IR based relevance matching module that measures keyword matches using n-gram embedding to generate an interaction matrix, and a semantic matching module that learns the semantic relationship between query and code by the co-attention mechanism. 

To evaluate the effectiveness of CSRS, we conduct a series of experiments on the dataset shared by Gu et al. \cite{b3} which contains 18.22M and 10k query-code pairs for the training set and testing set separately. We compare CSRS with two DL-based baseline models DeepCS and CARLCS-CNN. Experimental results show that CSRS achieves an MRR (mean reciprocal rank) of 0.614, outperforming DeepCS and CARLCS-CNN by 33.77\% and 18.53\% respectively.

In summary, this paper makes the following contributions:
\begin{itemize}
\item We introduce the neural IR method into code retrieval and design a relevance matching module based on neural IR that captures keyword matching signals (e.g. words, terms, and phrases) between query-code pairs.
\item We design a semantic matching module based on the co-attention mechanism which can learn the semantic representation of queries and code snippets.
\item We propose a novel code search model CSRS that combines our neural IR based relevance matching module and the semantic matching module. CSRS can simultaneously capture keyword matching signals and semantic correlations between queries and codes.
\item We evaluate CSRS on a large-scale dataset. The experimental results demonstrate that our model can advance code search tasks compared to two state-of-the-art models DeepCS and CARLCS-CNN.
\end{itemize}

The rest of the paper is organized as follows. Section \ref{section:RW} discusses the related work about code search and neural IR. Section \ref{section:PR} provides some background knowledge on code representation and attention mechanism. Then, we introduce the details of our proposed model CSRS in Section \ref{section:PM}. Section \ref{section:ES} and \ref{section:RS} describe the experiment settings, the evaluation results and give some illustrative examples to show the performance of CSRS. Section \ref{section:DSC} discusses threats to validity. Finally, Section \ref{section:CFW} concludes the paper and illustrates future work.

\section{Related Work} \label{section:RW}
The purpose of code search task is to search for relevant code snippets in the code base according to the query statements entered by the user, while the neural IR models are usually used to retrieve relevant documents, which is similar to code search. Thus, we will introduce the related work of code search (Sec.\ref{section:RWCS}) and neural IR (Sec.\ref{section:RWNIR}) in the next two subsections.

\subsection{Code Search} \label{section:RWCS}
In recent years, a lot of studies have emerged to improve the performance of code search. The prior studies are mostly based on information retrieval techniques, which focus on query reformulation and keyword matching between natural language queries and code snippets. For instance, McMillan et al. \cite{b8} proposed Portfolio, which returns a series of code fragments through keyword matching and PageRank. Haiduc et al. \cite{b9} proposed a machining learning model Refoqus to recommend a reformulation strategy based on the properties of the query to improve its performance. Sourcerer \cite{b12} is an infrastructure for large-scale code search based on Lucene\footnote{https://lucene.apache.org/}.

In recent years, with the rapid development of deep learning, many researchers use deep learning techniques to implement code search models. Gu et al. \cite{b3} first designed a deep learning model DeepCS, which uses the LSTMs to embed codes and queries into the unified semantic space. Based on DeepCS, Shuai et al. \cite{b4} proposed a co-attentive representation learning model to capture the correlations between query and code. Xu et al. \cite{b13} introduced a two-stage attention-based model to learn the semantic correlation effectively and efficiently.

In addition, many deep learning models use the structure information of the code. Wan et al. \cite{b14} developed a multimodal representation method to represent the unstructured and structured features of source code, in which LSTM, tree LSTM, and GGNN (Gating Graph Neural Network) \cite{b30} are used to represent the tokens, the AST (Abstract Syntax Tree) and the CFG (Control Flow Graph) of the code. Sun et al. \cite{b15} proposed PSCS that encodes both the semantics and structures of code represented by AST paths. Ling et al. \cite{b16} represented both queries and code snippets with the unified graph-structured data, and then match two graphs to retrieve the best code snippet.

Furthermore, some studies improve the performance of code search by introducing other code-related tasks. Yao et al. \cite{b17} proposed a code annotation model to generate code comments of the given code snippets that can be leveraged by the code retrieval model, which can distinguish relevant codes better from others. Ye et al. \cite{b18} utilized the code generation model to improve code retrieval via dual learning. Bui et al. \cite{b19} introduced a self-supervised contrastive learning framework to alleviate the need of labeled data for code retrieval.

In general, although DL-based models have achieved considerable results in the code search task, most of the above methods convert the query and code into a vector representation, which only retains the semantic information of queries and codes. While the IR-based models can only obtain the lexical correlations between queries and code snippets, which lack the support of semantic information. We believe that the combination of the two categories of models can enrich the matching information and improve the ability of the code search model.

\subsection{Neural Information Retrieval} \label{section:RWNIR}
In general, the aim of IR is to retrieve the relevant documents according to the queries. The IR methods represent queries and documents by words. The final ranking is based on the lexical matching between query words and document words, and the matching features are manually defined, which can be incomplete and time-consuming. Recently, with the successes of deep learning in many related areas, neural IR methods have the ability to improve the performance of traditional IR.

Existing neural IR models can be divided into two categories: representation based and interaction based \cite{b20}. The early studies mainly focus on representation based models, which try to construct representations of queries and documents, and the ranking is to match the similarity of the representations. For example, DSSM \cite{b21} projects queries and documents into a low-dimensional space where the distance is computed as the relevance between them. Based on the DSSM, Shen et al. \cite{b22} used a convolutional-pooling structure to improve the performance. These methods directly learn the representation of queries and documents without considering the interactions between them.

The interaction based models learn word-level matching signals by building the translation matrix from word pairs between queries and documents, then summarizing it into a ranking score. Guo et al. \cite{b20} utilized the pyramid pooling technique to summarize the translation matrix. Xiong et al. \cite{b23} proposed K-NRM that uses kernel-based pooling to obtain matching signals at different levels. Mitra et al. \cite{b24} combined representation based and interaction based models to improve the performance.

In the document retrieval task, compared with traditional IR methods, neural IR methods can capture lexical matching signals through neural networks without manually defining matching features. Inspired by this, we use the neural IR based method to construct the relevance matching module for queries and codes.

\section{Preliminaries} \label{section:PR}
In this section, we briefly introduce some background knowledge of CSRS. We first discuss the code representation using deep learning techniques for the code search task in subsection \ref{section:PCR}. Then we illustrate the preliminary of attention mechanism in subsection \ref{section:PAM}.

\subsection{Code Representation} \label{section:PCR}
In the code search task, code representation is a basic step toward understanding the code. Like the neural language, a code snippet should be mapped into a vector space so that it can be fed to the deep neural network for subsequent calculations. 

We can extract three features from the code snippet shown in Fig.\ref{fig:cseg}: (1) method name, a sequence of tokens split by camel case; (2) API sequence, a list of API words called in the code snippet; (3) code tokens, a list of words used in the method body. First, a vocabulary is employed to encode the words of the three features. Then the words in three features are converted into vectors with the same dimension. Each feature can be represented by a vector matrix. Finally, the three original vector matrices are fed into neural networks to obtain the embedded three feature matrices.

In the existing researches, DeepCS embeds the API sequence $\mathbf{s_{api}}$ and method name $\mathbf{s_{name}}$ by LSTM, while the code tokens $\mathbf{s_{tok}}$ are embedded by a multilayer perceptron (MLP)\cite{b33}. The final code representation $\mathbf{c}$ for DeepCS is computed in Eq.(\ref{equation:eq1}). As shown in Eq.(\ref{equation:eq2}), CARLCS-CNN uses LSTM to embed the API sequence, and the method name and code tokens are embedded by CNN. Recently, TabCS proposes to use the attention mechanism to encode three code features, as shown in Eq.(\ref{equation:eq3}).

\begin{equation} \label{equation:eq1}
\mathbf{c}=LSTM(\mathbf{s_{api}})+LSTM(\mathbf{s_{name}})+MLP(\mathbf{s_{tok}})
\end{equation}
\begin{equation} \label{equation:eq2}
\mathbf{c}=LSTM(\mathbf{s_{api}})+CNN(\mathbf{s_{name}})+CNN(\mathbf{s_{tok}})
\end{equation}
\begin{equation} \label{equation:eq3}
\mathbf{c}=Att(\mathbf{s_{api}})+Att(\mathbf{s_{name}})+Att(\mathbf{s_{tok}})
\end{equation}

\subsection{Attention Mechanism} \label{section:PAM}
Encoder-decoder models have achieved a lot of success in tasks like machine translation and text summarization. As the sentence length increases, the performance of the basic encoder-decoder model will decline sharply. To tackle this issue, the attention mechanism was proposed by Bahdanau et al. \cite{b27}, which allows the model to focus on the relevant parts of the long input sequence. The basic attention mechanism takes the input sequence as three matrices: query matrix ($Q$), key matrix ($K$), and value matrix ($V$), and each matrix consists of word vectors. The aim of the attention mechanism is to weigh the value vector by calculating the attention score through the query vector and key vector. Formally, the attention score can be computed as follows:
\begin{equation}
    Attention(Q,K)=g(f(Q,K))
\end{equation}
where $g$ is the activation function, and $f$ is the attention score function which can be a multi-layer perceptron \cite{b27}, dot product function \cite{b28}, and scaled dot product \cite{b29}. After the calculation of the attention score, we obtain the final attention by weighing the value vector using the attention scores:
\begin{equation}
    \alpha=\sum_i Attention_i(Q_i,K_i)V_i
\end{equation}
where $Attention_i$ is the $i$-th attention score, and $Q_i$, $K_i$, $V_i$ are the $i$-th element in query, key and value vector matrix respectively.

\section{Proposed Model} \label{section:PM}
\begin{figure*}[htbp]
    \centering
	\includegraphics[width=1.0\textwidth]{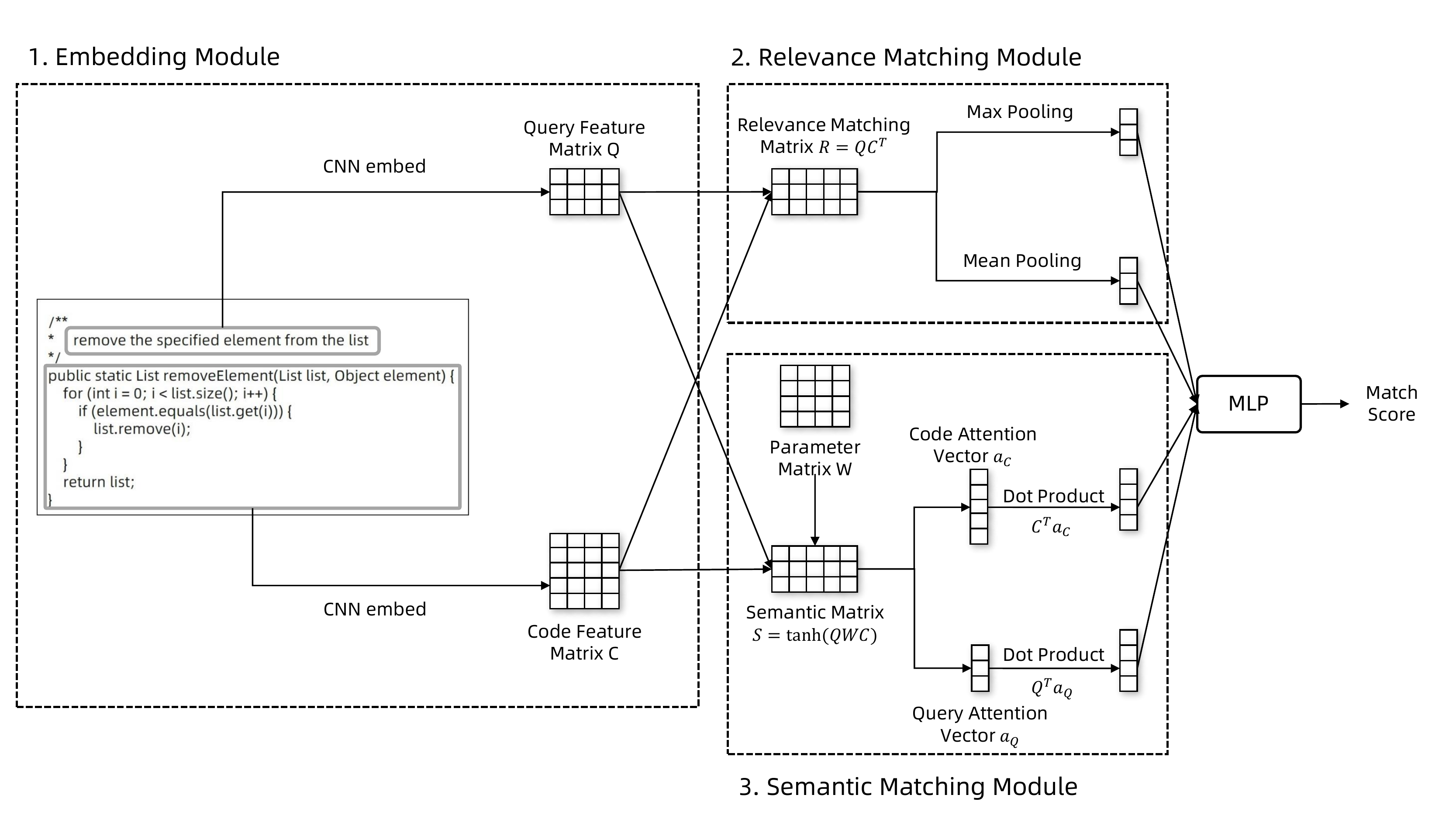}
	\caption{The architecture of CSRS. The model consists of three major modules: (1) a CNN-based embedding module that generates n-gram embeddings; (2) a relevance matching module for learning keyword matching signals; (3) a semantic matching module with co-attention mechanisms for learning semantic matching signals.}
	\label{fig:arch}
\end{figure*}

In this section, we present the design and implementation details of our proposed model CSRS.

\subsection{Overview}
Fig.\ref{fig:arch} illustrates the overall structure of CSRS. The model is composed of three major parts: (1) a CNN based embedding module, which uses convolution kernels of different sizes to generate n-gram embeddings of descriptions and codes (Sec.\ref{section:EM}); (2) a relevance matching module that measures keyword matches (e.g. words, terms, and phrases) between descriptions and code snippets (Sec.\ref{section:RM}); (3) a semantic matching module with the co-attention mechanism for learning semantic matching signals (Sec.\ref{section:SM}). Finally, we feed the relevance matching features and semantic matching features to an MLP, and then output the matching score of the description and code.

\subsection{Embedding Module} \label{section:EM}
\textbf{Code Embedding.} Each code snippet consists of three parts: method name, API sequence, and tokens. Firstly, for each feature of the code, we use an embedding layer to convert each word into a vector. However, using this kind of embedding is equivalent to treating each word as a unigram. In this way, it is difficult to express the information of specific terms composed of multiple words in the code, such as "quickSort", "HashMap" and so on. We hope to match these terms as n-grams instead of splitting them into unigrams for matching. Therefore, we employ CNN to compose adjacent words’ embeddings to n-gram embeddings. We implement the n-gram embeddings of description and code through the following steps.

\begin{figure}[htp]
    \centering
	\includegraphics[width=0.48\textwidth]{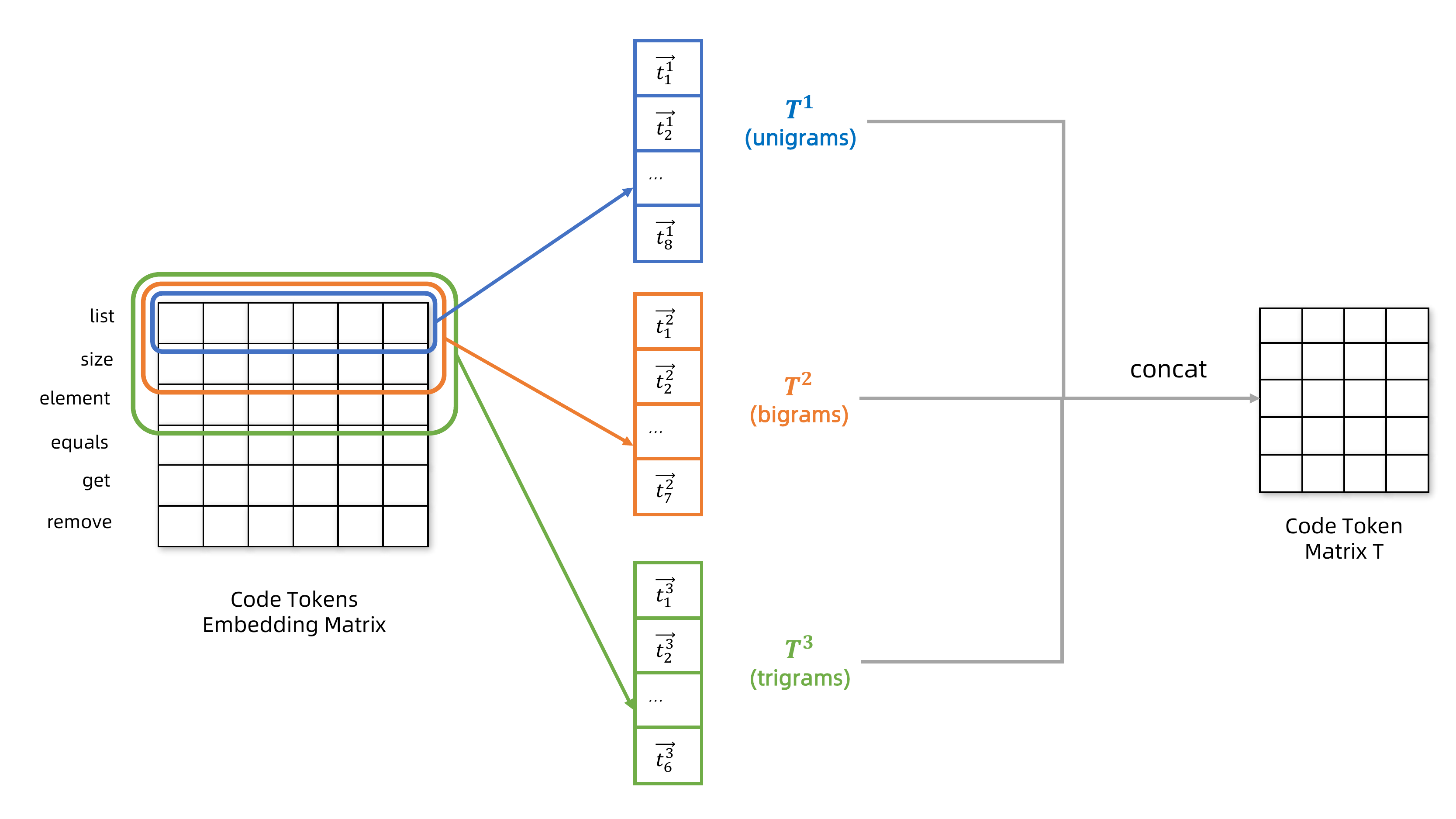}
	\caption{The embedding process of code tokens.}
	\label{fig:emb}
\end{figure}

Firstly, code token is a list of words extracted from the method body, in which duplicate words, stop words, and Java keywords are not included. 

The implementation of n-gram embeddings for code tokens is shown in Fig.\ref{fig:emb}. Let $\mathbf{e^T_i}\in\mathbb{R}^d$ be a $d$-dimensional word initial vector corresponding to the $i$-th word in code tokens. A code tokens sequence with length $n_t$ can be represented by an embedding matrix $E^T\in\mathbb{R}^{n_t\times d}$ as shown in Eq.(\ref{equation:eq6}).
\begin{equation}
    E^T=[\mathbf{e^T_1},\mathbf{e^T_2},\dots,\mathbf{e^T_{n_t}}] \label{equation:eq6}
\end{equation}
Then the convolutional layer applies convolution filters to compose n-grams from the code tokens. We use $d$ different convolution filters $W^T_1,\dots,W^T_d\in\mathbb{R}^{h \times d}$ slide over the code tokens embedding matrix $E^T$ like a sliding window, where $h$ from 1 to 3 denotes the width of convolution kernels. For each window of $h$ words, the filter $W^T_j$ sums up all elements in the $h$ words’ embeddings $E^T_{i:i+h-1}$, and produces a feature score $v^t_j$ as follows:
\begin{equation}
    v^t_j=f(W^T_j*E^T_{i:i+h-1}+\mathbf{b})
\end{equation}
where $j\in \{1,2,\dots,d\}$, $\mathbf{b}\in\mathbb{R}$ is a bias term, $*$ is the convolution operator and $f$ is a non-linear function such as the hyperbolic tangent. The $d$ filters produce $d$ feature scores, so that we can obtain a $d$-dimensional code tokens embedding for the h-gram:
\begin{equation}
    \mathbf{t^h_i}=[v^t_1,v^t_2,\dots,v^t_d]
\end{equation}
Thus, the convolution layer converts the code tokens embedding matrix $E^T$ into h-gram embedding matrix $T^h\in\mathbb{R}^{(n_t-h+1)\times d}$.
\begin{equation}
    T^h=[\mathbf{t^h_1},\mathbf{t^h_2},\dots,\mathbf{t^h_{n_t}}]
\end{equation}

Specifically, the window size $h$ ranges from 1 to 3, which means we can get three $h$-gram embedding matrices: $T^1$ for unigram, $T^2$ for bigram, and $T^3$ for trigram. Then, three h-gram embedding matrices are concatenated into the final code tokens matrix $T$:
\begin{equation}
    T=T^1 \oplus T^2 \oplus T^3
\end{equation}
where $\oplus$ is the concatenation operator.

For a given method name, such as "getValue", we split it into a sequence of words according to the camal-case naming convention. Given a method name sequence of length $n_m$, its original embedding matrix is $E^M=[\mathbf{e^M_1},\dots,\mathbf{e^M_{n_m}}]\in\mathbb{R}^{n_m\times d}$, where $\mathbf{e^M_i}$ denotes the $i$-th word embedding in method name sequence. The n-gram embedding matrix $M$ for the method name are generated using the same way as code token embedding.

Given an API sequence of length $n_a$, such as "getThemeImage, Map, put", its original embedding matrix is $E^A=[\mathbf{e^A_1},\dots,\mathbf{e^A_{n_a}}]\in\mathbb{R}^{n_a\times d}$, where $\mathbf{e^A_i}$ denotes the $i$-th word embedding in the API sequence. We use the same method to obtain the final API feature matrix $A$.

After converting the three parts of the code into n-gram embeddings, we get three code feature matrices, we eventually concatenate them into the final code feature matrix $C\in\mathbb{R}^{n\times d}$.
\begin{equation}
    C=T \oplus M \oplus A
\end{equation}

\textbf{Description Embedding.} The description indicates the user's query intention, which also contains special terms. Therefore, for the given description $E^D=[\mathbf{e^D_1},\dots,\mathbf{e^D_m}]\in\mathbb{R}^{m\times d}$, where $\mathbf{e^D_i}$ denotes the $i$-th word embedding in the description. As shown in Eq.(\ref{equation:eq12})-(\ref{equation:eq13}), we use $d$ different convolution kernels $W^D_1,\dots,W^D_d\in\mathbb{R}^{h \times d}$ for convolution operation, and the feature scores $v^d_j$ generated by each convolution kernel are considered as n-gram embedding vectors $d^h_i$.
\begin{equation}
    v^d_j=f(W^D_j*E^D_{i:i+h-1}+\mathbf{b}) \label{equation:eq12}
\end{equation}
\begin{equation}
    \mathbf{d^h_i}=[v^d_1,v^d_2,\dots,v^d_d] \label{equation:eq13}
\end{equation}
The h-gram embedding matrix $D^h$ consists of n-gram embedding vectors $d^h_i$. We obtain the final description feature matrix $D\in\mathbb{R}^{m\times d}$ by concatenating $h$ n-gram embedding matrices:
\begin{equation}
    D^h=[\mathbf{d^h_1},\mathbf{d^h_2},\dots,\mathbf{d^h_m}]
\end{equation}
\begin{equation}
    D=D^1 \oplus D^2 \oplus D^3
\end{equation}

\subsection{Relevance Matching Module} \label{section:RM}
This section describes our efforts to capture keyword matching signals for relevance matching, which measures soft term matches between description-query pairs. We build the interaction matrix $R$ by multiplying the description feature matrix $D\in\mathbb{R}^{m\times d}$ and the code feature matrix $C\in\mathbb{R}^{n\times d}$, which aims to calculate the relevance score between the description and the code:
\begin{equation}
    R=DC^T
\end{equation}
where $R_{i,j}$ can be considered the similarity score by matching the description n-gram vector $D[i]$ with the code n-gram vector $C[j]$. 

Since in the description and code, similar n-grams will have closer embedding vectors, their product will produce larger scores. Next, we obtain a normalized relevance matching matrix $\hat{R}$ by applying a softmax function over the code columns of $R$ to normalize the similarity scores into the $[0,1]$ range. For each description n-gram $i$, the softmax function normalizes its matching scores over all n-grams in the code and makes the similar n-grams have a score closer to 1.0. Then we leverage $max$ and $mean$ pooling to obtain two relevance matching feature vectors for the matrix $\hat{R}$:
\begin{equation}
    \mathbf{o^{RM}_{max}}=[max(\hat{R}_{1,:}),\dots,max(\hat{R}_{n,:})]
\end{equation}
\begin{equation}
    \mathbf{o^{RM}_{mean}}=[mean(\hat{R}_{1,:}),\dots,mean(\hat{R}_{n,:})]
\end{equation}

\subsection{Semantic Matching Module} \label{section:SM}
In addition to relevance matching, the semantic matching module aims to capture semantic correlations via co-attention mechanisms on the description and code feature matrices. Unlike the basic attention mechanism, the co-attention mechanism can focus on the description attention and the code attention simultaneously, and learn the semantic vector representation of both.

In detail, we take the description feature matrix $D\in\mathbb{R}^{m\times d}$ and code feature matrix $C\in\mathbb{R}^{n\times d}$ as the query matrix and the key matrix of the attention mechanism. The attention matrix $S\in\mathbb{R}^{m\times n}$ can be calculated by $D$, $C$ and a parameter matrix $W\in\mathbb{R}^{d\times d}$ learned by the neural networks. We use the $tanh$ activation function to scale each element in $S$ between -1 and 1.
\begin{equation}
    S=tanh(DWC)
\end{equation}

The attention matrix $S$ can focus on the semantic correlation between descriptions and codes. $S_{i,j}$ represents the semantic matching score between the description n-gram feature vector $D[i]$ and the code n-gram feature vector $C[j]$. Specifically, the $i$-th row in $S$ denotes the semantic correlations of the $i$-th n-gram in description to each n-gram in code. Similarly, the $j$-th column in $S$ denotes the semantic correlations of the $j$-th n-gram in code to each n-gram in the description.

Then, we employ $max$ pooling along rows and columns over $S$ as follows, which denotes that we can focus on the description attention and the code attention:
\begin{equation}
    \mathbf{u}^D=[max(S_{1,:}),\dots,max(S_{m,:})]
\end{equation}
\begin{equation}
    \mathbf{u}^C=[max(S_{:,1}),\dots,max(S_{:,n})]
\end{equation}
where the $i$-th element of $\mathbf{u}^D\in\mathbb{R}^m$ represents a semantic score between the $i$-th n-gram in description $D$ and its most semantically similar n-gram in code $C$, and the $i$-th element of $\mathbf{u}^C\in\mathbb{R}^n$ represents a semantic score between the $i$-th n-gram in code $C$ and its most semantically similar n-gram in description $D$. Next, we obtain the description attention weight vector and the code attention weight vector by transforming $\mathbf{u}^D$ and $\mathbf{u}^C$ into $\mathbf{a}^D$ and $\mathbf{a}^C$ using softmax function.
\begin{equation}
    \mathbf{a}^D_i=\frac{\exp{(\mathbf{u}^D)}}{\sum^m_{p=1}\exp{(\mathbf{u}^D_p)}}, 
    \quad\mathbf{a}^C_i=\frac{\exp{(\mathbf{u}^C)}}{\sum^n_{q=1}\exp{(\mathbf{u}^C_q)}}
\end{equation}
\begin{equation}
    \mathbf{a}^D=[\mathbf{a}^D_1,\dots,\mathbf{a}^D_m], 
    \quad\mathbf{a}^C=[\mathbf{a}^C_1,\dots,\mathbf{a}^C_n]
\end{equation}

Finally, based on the two attention weight vectors, the final description attention vector $\mathbf{o}^{SM}_{desc}$ and the code attention vector $\mathbf{o}^{SM}_{code}$ can be calculated as the weighted sum of the description feature matrix $D$ and the code feature matrix $C$. We take the two vectors $\mathbf{o}^{SM}_{desc}$ and $\mathbf{o}^{SM}_{code}$ as the description semantic vector and the code semantic vector.
\begin{equation}
    \mathbf{o}^{SM}_{desc}=D^T\mathbf{a}^D, 
    \quad\mathbf{o}^{SM}_{code}=C^T\mathbf{a}^C
\end{equation}

\subsection{Prediction}
After all the computation of the relevance matching module and the semantic matching module, we obtain the matching features $\{\mathbf{o^{RM}_{max}},\mathbf{o^{RM}_{mean}},\mathbf{o}^{SM}_{desc},\mathbf{o}^{SM}_{code}\}$. Then we concatenate all features to the final feature vector $\mathbf{o}$, and feed $\mathbf{o}$ to a two-layer perceptron followed by a ReLU activation. We express the output of the two-layer perceptron as the matching score, and divide the scores into two categories. The first class indicates that the input natural language description is related to the code, and the output value is closer to 1, while the second class indicates that the input natural language description is not related to the code, and the output value is closer to 0.
\begin{equation}
    \mathbf{o}=\mathbf{o^{RM}_{max}}\otimes\mathbf{o^{RM}_{mean}}\otimes\mathbf{o}^{SM}_{desc}\otimes\mathbf{o}^{SM}_{code}
\end{equation}

\subsection{Training}
During training, we let each training data be a triple $<\mathcal{D},\mathcal{C},\mathcal{L}>$, where $\mathcal{D}$ is the description, $\mathcal{C}$ is the code and $\mathcal{L}$ represents the ground truth class. We use Adam algorithm \cite{b10} to minimizing cross-entropy loss in the training process:
\begin{equation}
    loss=-\sum^2_{i=1}l(i)\log g(i)
\end{equation}
where $l$ denotes the ground truth class, $g$ is the matching score computed by CSRS.

\subsection{Implementation Details}
To conduct our experiment, we set the word embedding size to 100, and the batch size is set as 128. In order to avoid overfitting, we use the dropout technique \cite{b31} and set the drop rate to 0.25. We initialize the Adam optimizer \cite{b10} with the learning rate 0.0001. All the experiments are implemented using the Keras 2.3.1 framework with Python 3.6, and the experiments are conducted on a computer with a GeForce RTX 2080 Ti GPU with 11 GB memory, running on Ubuntu 16.04.

\section{Experiment Setup} \label{section:ES}
In this section, we first report several research questions (RQs) and experiment results. Then we describe the dataset, evaluation metrics, and several baselines.
\subsection{Research Questions} \label{section:RQ}
\noindent Our work aims to answer the following four research questions:
\begin{itemize}
    \item \textbf{RQ1: How does our proposed model perform?}
    
    To answer this question, we conduct an experiment to investigate whether our proposed model performs better than two state-of-the-art baseline models DeepCS \cite{b3} and CARLCS-CNN \cite{b4}.
    
    \item \textbf{RQ2: What is the contribution of each module in CSRS?}
    
    In RQ2, we focus on exploring the contribution of the relevance matching module and semantic matching module to CSRS. Specifically, we start with the complete model and then remove each module in turn to understand its effectiveness.
    
    \item \textbf{RQ3: How do the three code features affect the model?}
    
    We use three code features (i.e. method name, API sequence, and code tokens) to represent a whole code method. In order to understand the impact of these three code features on CSRS, we train CSRS using individual features separately, and then compare with the complete model to analyze the impact of each feature.
    
    \item \textbf{RQ4: How does the window size affect the model effectiveness?}
    
    In the embedding module, we let the window size be 1, 2, and 3 to capture unigram, bigram, and trigram respectively. This RQ mainly focuses on the choice of window size. Due to analyzing the effectiveness of the choice of window size, we run CSRS with individual window size so that the model can only generate one n-gram embedding.
    
\end{itemize}

\subsection{Dataset}
We evaluate our approach on Gu's \cite{b3} dataset. The dataset for training contains 18,223,872 commented Java code methods collected from GitHub repositories with at least one star from August 2008 to June 2016. The testing dataset contains 10,000 code-query pairs. As shown in Table \ref{table:t1}, we also count the average tokens of the description and the three features of code snippets in the training and testing dataset.
\begin{table}[]
    \caption{Statistics For Java-small Dataset}
	\begin{center}
        \begin{tabular}{cccccc}
        \hline
          \textbf{Dataset} &
          \textbf{Size} &
          \begin{tabular}[c]{@{}c@{}}\textbf{Avg}\\ \textbf{Query}\end{tabular} &
          \begin{tabular}[c]{@{}c@{}}\textbf{Avg}\\ \textbf{Token}\end{tabular} &
          \begin{tabular}[c]{@{}c@{}}\textbf{Avg}\\ \textbf{Methname}\end{tabular} &
          \begin{tabular}[c]{@{}c@{}}\textbf{Avg}\\ \textbf{Apiseq}\end{tabular} \\ \hline
        Training &
          18223872 &
          9.89 &
          10.30 &
          2.48 &
          8.39 \\
        Testing &
          10000 &
          9.87 &
          10.29 &
          2.49 &
          8.33 \\ \hline
        \end{tabular}
    \end{center}
    \label{table:t1}
\end{table}

\subsection{Evaluation Metrics}
We evaluate the performance of CSRS by using three common metrics: recall, MRR (Mean Reciprocal Rank) and NDCG (Normalized Discounted Cumulative Gain) \cite{b32}. The three metrics are defined as follows:

\begin{itemize}
    \item \textbf{Recall@k} calculates the proportion of queries that the relevant code could be found in the top-$k$ ranked list. $Recall@k$ is computed as follows:
    \begin{equation}
        Recall@k=\frac{1}{|\mathcal{Q}|}\sum^{|\mathcal{Q}|}_{i=1}I(\mathcal{Q}_i \leq k)
    \end{equation}
    where $\mathcal{Q}$ is the 10,000 queries in testing dataset, $I$ is the indicator function that returns 1 if the code related to $i$-th query $\mathcal{Q}_i$ is in the top-k list, otherwise it returns 0. We evaluate $Recall@k$ when $k$'s value equals 1, 5 and 10. These values reflect the typical lengths of ranked list that users would focus\cite{b11}.
    \item \textbf{MRR} counts the average of the reciprocal ranks of all queries $\mathcal{Q}$. The MRR is computed as follows:
    \begin{equation}
        MRR=\frac{1}{|\mathcal{Q}|}\sum^{|\mathcal{Q}|}_{i=1}\frac{1}{Rank_{\mathcal{Q}_i}}
    \end{equation}
    where $\mathcal{Q}$ is the 10,000 queries in testing dataset, $Rank_{\mathcal{Q}_i}$ denotes the rank of the ground truth code related to the $i$-th query $\mathcal{Q}_i$ in the top-$k$ ranked list. Following Gu et al.\cite{b3}, We let $k$ equal to 10. If the ground truth code is not in the ranked list, we set the $\frac{1}{Rank_{\mathcal{Q}_i}}$ equal to 0.
    \item \textbf{NDCG} measures the correlation between the ranking result of code snippets and the query. The idea of $NDCG$ is that the results of high correlation affect the final $NDCG$ score more than the results of general correlation, and when the results with high correlation appear in the front position, the $NDCG$ score will be higher. We calculate $NDCG$ as follows:
    \begin{equation}
        NDCG=\frac{1}{|\mathcal{Q}|}\sum^k_{i=1}\frac{2^{r_i}-1}{\log_2(i+1)}
    \end{equation}
    where $\mathcal{Q}$ denotes the 10k queries in testing dataset, $k$ equals to 10, and $r_i$ is the relevance score of the top-$k$ results at position $i$.
\end{itemize}

\subsection{Baselines}
\noindent\textbf{DeepCS}. The first model uses the deep learning method proposed by Gu et al. \cite{b3}, which embeds query and code features separately with two different LSTM models. We reproduce the DeepCS by the open source code shared on GitHub\footnote{https://github.com/guxd/deep-code-search}.

\noindent\textbf{CARLCS-CNN}. A code search model based on CNN, which uses co-attention mechanism to construct the semantic relationship between code snippets and their queries. We also reproduce the source code shared by Shuai et al.\footnote{https://github.com/cqu-isse/CARLCS-CNN} \cite{b4}.

\section{Results} \label{section:RS}
In this section, we first report our experimental results and answer research questions described in Sec \ref{section:RQ} respectively. Then, we provide several illustrative examples to show the effect of CSRS.
\subsection{RQ1: Model Effectiveness}
We compare CSRS with two state-of-the-art models DeepCS and CARLCS-CNN under the $Recall@k$, $MRR$ and $NDCG$ evaluation metrics. The evaluation results are presented in Table \ref{table:t2}.
\begin{table}[htbp]
    \caption{Comparison of the effects of DeepCS and CARLCS-CNN}
	\begin{center}
	\setlength{\tabcolsep}{0.8mm}{
	    \begin{tabular}{cccccc}
        \hline
        \textbf{Model}      & \textbf{Recall@1} & \textbf{Recall@5} & \textbf{Recall@10} & \textbf{MRR} & \textbf{NDCG}   \\ \hline
        DeepCS     & 0.337 & 0.624 & 0.720 & 0.459 & 0.522 \\
        CARLCS-CNN & 0.402 & 0.674 & 0.762 & 0.518 & 0.577 \\ \hline
        CSRS & \textbf{0.486} & \textbf{0.787} & \textbf{0.864} & \textbf{0.614} & \textbf{0.674} \\ \hline
        \end{tabular}
    }
    \end{center}
    \label{table:t2}
\end{table}

As we can see from Table \ref{table:t2}, CSRS provides the best results compared with two state-of-the-art models. For the metric of 
$Recall$
$@1/5/10$, CSRS achieves 0.486/0.787/0.864 respectively, which outperforms the DeepCS and CARLCS-CNN by 44.21\%/26.12\%/20.00\% and 20.90\%/16.77\%/13.39\% respectively. CSRS achieves an $MRR$ of 0.614, which improves DeepCS by 33.77\% and CARLCS-CNN by 18.53\%. In addition, CSRS achieves 0.674 for $NDCG$, which advances DeepCS and CARLCS-CNN by 29.12\% and 16.81\%.

\begin{tcolorbox}
\textbf{Result 1:} Our proposed model outperforms two state-of-the-art models DeepCS and CARLCS-CNN with the evaluation metrics of $Recall@k$, $MRR$ and $NDCG$.
\end{tcolorbox}

\subsection{RQ2: The contribution of Two Matching Modules}
To answer RQ2, we conduct the ablation experiment to investigate the contribution of two matching modules. We test by removing one matching module separately:

\textbf{Relevance Matching (RM)}: The relevance matching module focus on the lexical matching between the queries and code snippets. We remove the semantic matching module to evaluate the contribution of the relevance matching module.

\textbf{Semantic Matching (SM)}: The semantic matching module aims to capture the semantic correlation between the queries and code snippets. Similarly, we delete the relevance matching module to figure out the contribution of the semantic matching module.

\begin{table}[htbp]
    \caption{The contribution of each matching module}
    \begin{center}
    \setlength{\tabcolsep}{1mm}{
        \begin{tabular}{cccccc}
        \hline
        \textbf{Model}     & \textbf{Recall@1} & \textbf{Recall@5} & \textbf{Recall@10} & \textbf{MRR} & \textbf{NDCG}    \\ \hline
        CSRS(RM)   & 0.416             & 0.706    & 0.802     & 0.539     & 0.602  \\ 
        CSRS(SM)   & 0.452             & 0.737    & 0.823     & 0.574     & 0.634  \\ \hline
        CSRS & \textbf{0.486}             & \textbf{0.787}    & \textbf{0.864}     & \textbf{0.614}     & \textbf{0.674} \\ \hline
        \end{tabular}
    }
    \end{center}
    \label{table:t3}
\end{table}

Table \ref{table:t3} illustrates the reduction in metrics when the module is removed. \textbf{RM} results in a decrease of $MRR$ by 13.91\%, where the $MRR$ of \textbf{SM} drops by 6.97\%. For the metric $NDCG$, \textbf{RM} and \textbf{SM} have the decrease by 11.96\% and 6.31\%. In addition, for the $Recall@k$ evaluation metric, both \textbf{RM} and \textbf{SM} have different degrees of decline. The ablation experiment shows that the complete model is superior to \textbf{RM} and \textbf{SM}, which indicates the effectiveness of the combination of two matching modules.

\begin{tcolorbox}
\textbf{Result 2:} Both relevance matching and semantic matching are effective, while the combination of two matching modules performs the best.
\end{tcolorbox}

\subsection{RQ3: The Impact of Different Code Features}
Three code features are used in CSRS, including method name (M), API sequence (A), and code tokens (T). In order to find out the impact of these code features, we trained three models, each of which uses only M, A, or T as the input code feature. 

Table \ref{table:t4} reported the evaluation results. We can observe that only using the method name, API sequence or code tokens results in a drastic decrease of $MRR$ and $NDCG$, which denotes that all code features can improve the performance of CSRS. 

\begin{table}[htbp]
    \caption{Comparison of the effects of the three code features}
    \begin{center}
    \setlength{\tabcolsep}{1mm}{
        \begin{tabular}{cccccc}
        \hline
        \textbf{Model}     & \textbf{Recall@1} & \textbf{Recall@5} & \textbf{Recall@10} & \textbf{MRR} & \textbf{NDCG}   \\ \hline
        CSRS(M)         & 0.350             & 0.596    & 0.694     & 0.456     & 0.513 \\
        CSRS(A)         & 0.109             & 0.302    & 0.411     & 0.193     & 0.244 \\
        CSRS(T)         & 0.291             & 0.568    & 0.681     & 0.409     & 0.474 \\ \hline
        CSRS & \textbf{0.486}             & \textbf{0.787}    & \textbf{0.864}     & \textbf{0.614}     & \textbf{0.674} \\ \hline
        \end{tabular}
    }
    \end{center}
    \label{table:t4}
\end{table}

\begin{tcolorbox}
\textbf{Result 3:} The result shows that the three code features(i.e. method name, API sequence, and code tokens) have an effect on the proposed model CSRS, while the method name is the most effective feature.
\end{tcolorbox}

\subsection{RQ4: The Effects of Different Convolution Kernel Size in CNN}
In the embedding module, we use three different sizes of convolution kernels to obtain n-gram embeddings. To explore the impact of using only one size of kernel on the model, we ran three models which only use kernels with sizes of 1 (Conv1), 2 (Conv2), or 3 (Conv3) respectively.

\begin{table}[htbp]
    \caption{The effect of different sizes of convolution kernels on the effectiveness of CSRS}
    \begin{center}
    \setlength{\tabcolsep}{0.8mm}{
        \begin{tabular}{cccccc}
        \hline
        \textbf{Model}     & \textbf{Recall@1} & \textbf{Recall@5} & \textbf{Recall@10} & \textbf{MRR} & \textbf{NDCG}   \\ \hline
        CSRS(Conv1)     & 0.473             & 0.751    & 0.838     & 0.593     & 0.652 \\
        CSRS(Conv2)     & 0.386             & 0.699    & 0.799     & 0.518     & 0.585 \\
        CSRS(Conv3)     & 0.427             & 0.728    & 0.819     & 0.554     & 0.618 \\ \hline
        CSRS & \textbf{0.486}             & \textbf{0.787}    & \textbf{0.864}     & \textbf{0.614}     & \textbf{0.674} \\ \hline
        \end{tabular}
    }
    \end{center}
    \label{table:t5}
\end{table}

Table \ref{table:t5} shows that three different sizes of convolution kernels can improve the effectiveness of CSRS. As we expected, the kernel of size 1 contributes more to the complete model. Meanwhile, the best performance can be obtained by using three sizes of convolution kernels at the same time.

\begin{tcolorbox}
\textbf{Result 4:} The three different sizes of convolution kernels are all necessary. The convolution kernel with width 1 is close to the effect of the complete model, while the combination achieves the best performance.
\end{tcolorbox}

\subsection{Illustrative Examples}
\begin{figure}[htbp]
    \centering
    \subfigure[The top 1 result of CSRS]{
        \label{fig:cs1}
        \includegraphics[width=0.48\textwidth]{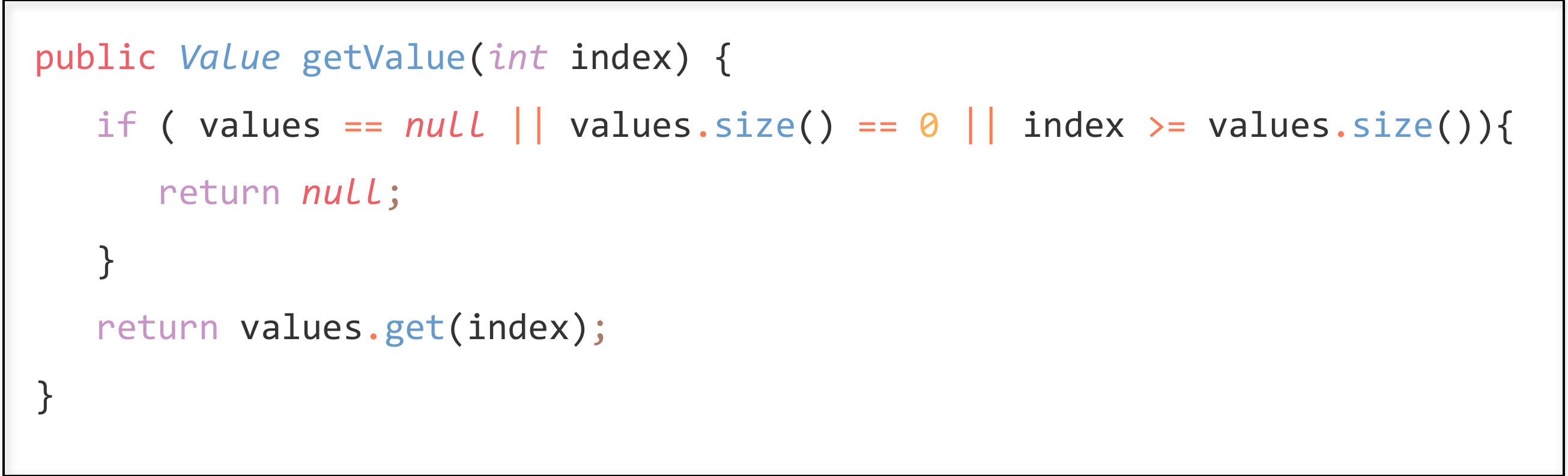}}
    \subfigure[The top 1 result of CARLCS-CNN]{
        \label{fig:cs2}
        \includegraphics[width=0.48\textwidth]{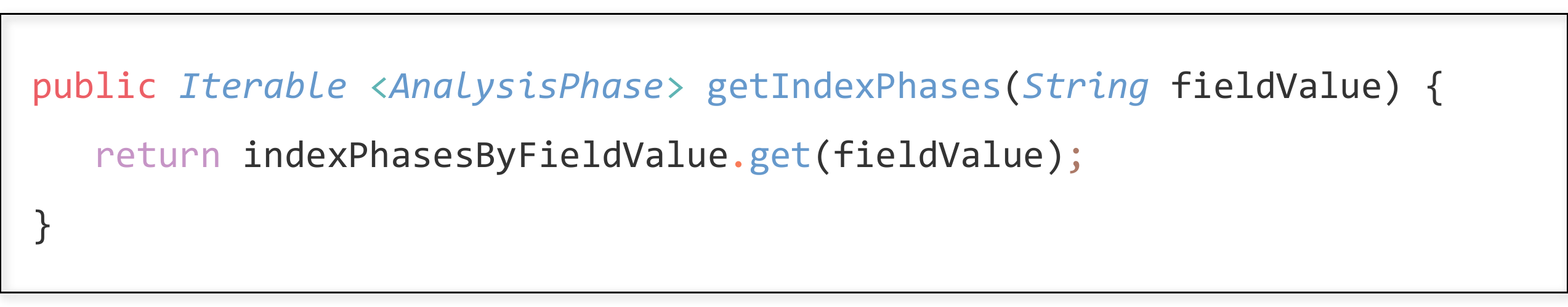}}
    \caption{The top 1 search result of CSRS and CARLCS-CNN for the query "return the value at index or null if that is out of range"}
    \label{fig:csmain}
\end{figure}

Fig.\ref{fig:csmain} shows the top 1 search result of CSRS and CARLCS-CNN for the query "return the value at index or null if that is out of range". It can be observed from Fig.\ref{fig:cs1} that CSRS can retrieve the ground truth code snippet, because there are many overlaps between the query and the code, such as "value", "index" and "null", which demonstrates the effectiveness of the relevance matching module in CSRS. Meanwhile, it also illustrates that CSRS can understand the semantic of "out of range" in the query. In contrast, Fig.\ref{fig:cs2} shows that CARLCS-CNN returns a irrelevant code.

\begin{figure}[htbp]
    \centering
    \subfigure[The top 1 result of CSRS]{
        \label{fig:cs3}
        \includegraphics[width=0.48\textwidth]{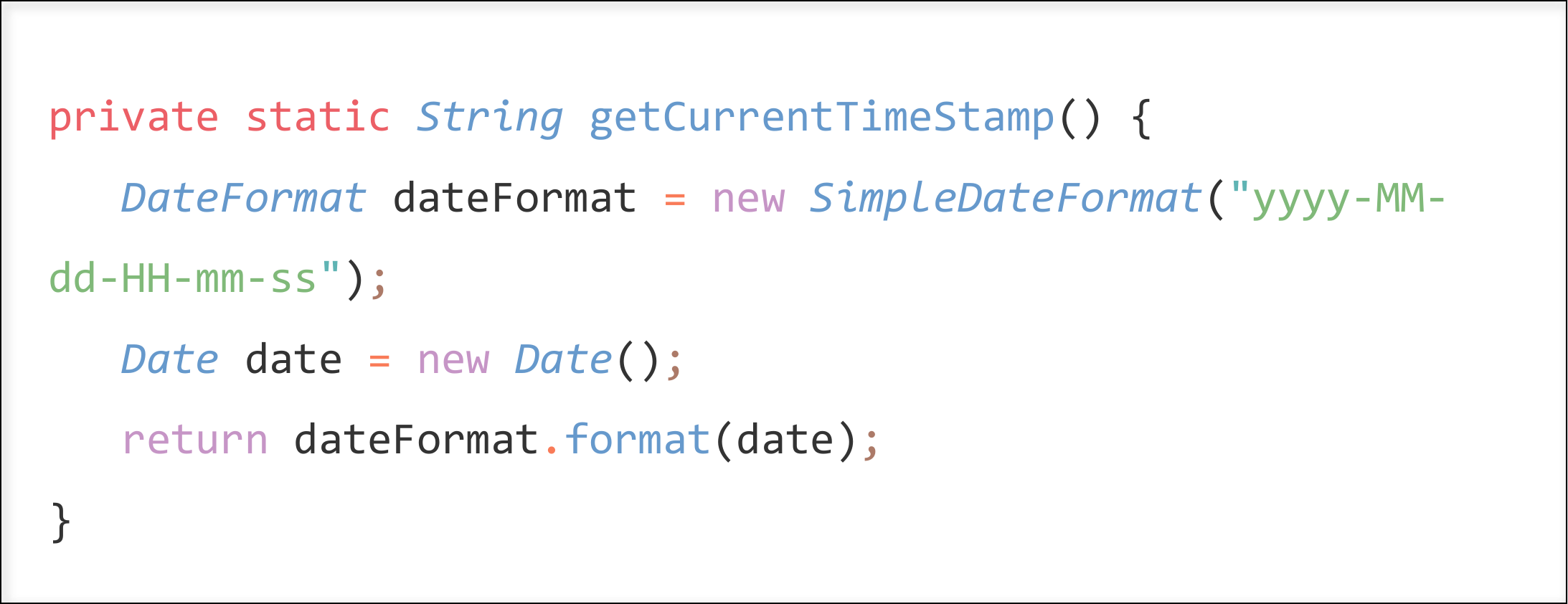}}
    \subfigure[The top 1 result of CARLCS-CNN]{
        \label{fig:cs4}
        \includegraphics[width=0.48\textwidth]{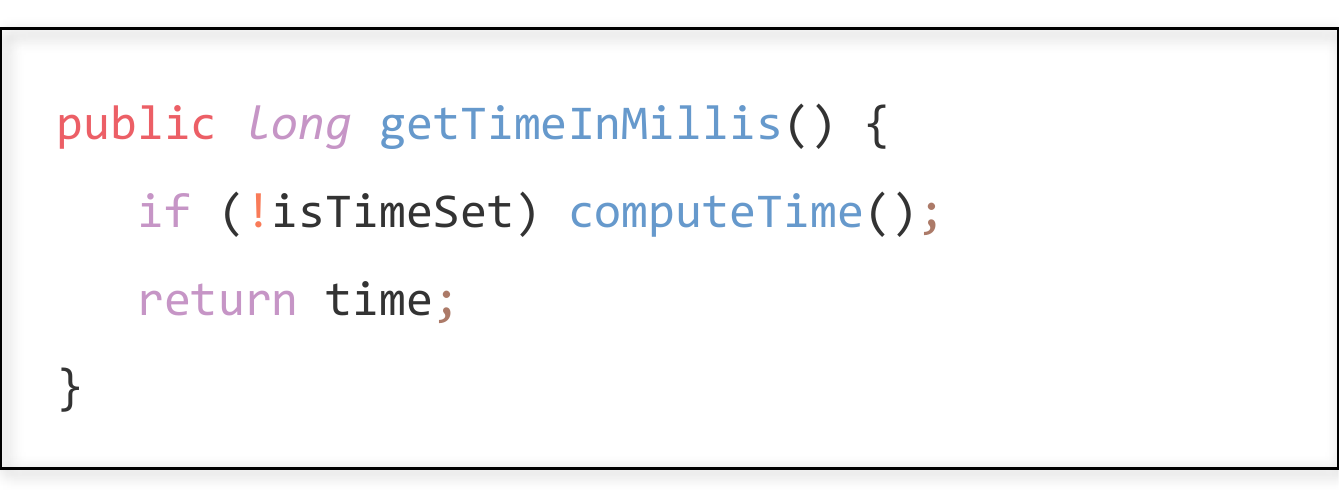}}
    \caption{Another search result of CSRS and CARLCS-CNN for the query "get current time stamp"}
    \label{fig:csmain2}
\end{figure}

Fig.\ref{fig:csmain2} shows another search result for the query "get current time stamp". Obviously, from Fig.\ref{fig:cs3}, we can see that the query words can match the method name exactly in the first result of CSRS. Moreover, the query word "time" semantically correlates with the word "date" in the code snippet. However, Fig.\ref{fig:cs4} shows that CARLCS-CNN can not retrieve the correct code.

To sum up, the above two cases illustrate that CSRS can capture both keyword matching information and semantic matching information.

\section{Threats to Validity} \label{section:DSC}
Our proposed model may suffer from two threats to validity. The first one is the implementation of baselines, reproducing the baseline models by ourselves may cause errors. To reduce this threat, we re-ran DeepCS and CARLCS-CNN with the source code and datasets shared by the authors in GitHub.

The second threat to this work is the generalization of the proposed approach. The dataset provided by Gu et al. \cite{b3} is obtained from the open source Java project on GitHub and contains only about 10,000 testing data. Therefore, the experimental results and conclusions may be different for using other programming language datasets or larger testing sets.

\section{Conclusion and Future Work} \label{section:CFW}
In this paper, we propose a novel deep learning model for code search. Our approach consists of two parts, a relevance matching module to capture n-gram matching signals and a semantic matching module for capturing the semantic correlation between the queries and the codes. Our experimental results show that CSRS outperforms DeepCS and CARLCS-CNN in terms of $MRR$ by 33.77\% and 18.53\%, and $NDCG$ by 29.12\% and 16.81\%, which achieves a significant improvement.

We consider two future works. First, we will try to design a more effective method to combine the relevance matching module and the semantic matching module. Second, we plan to incorporate external knowledge such as knowledge graphs or API documentation to enhance the quality of training data.

\section*{Acknowledgment}
This work has been supported by the Foundation item: National Key R\&D Program of China (2018YFB1402800); National Natural Science Foundation of China (61772560).

\balance
\normalem\bibliographystyle{ACM-Reference-Format}
\bibliography{main}


\begin{thebibliography}{33}

%%% the Web address.

\ifx \showCODEN    \undefined \def \showCODEN     #1{\unskip}     \fi
\ifx \showDOI      \undefined \def \showDOI       #1{#1}\fi
\ifx \showISBNx    \undefined \def \showISBNx     #1{\unskip}     \fi
\ifx \showISBNxiii \undefined \def \showISBNxiii  #1{\unskip}     \fi
\ifx \showISSN     \undefined \def \showISSN      #1{\unskip}     \fi
\ifx \showLCCN     \undefined \def \showLCCN      #1{\unskip}     \fi
\ifx \shownote     \undefined \def \shownote      #1{#1}          \fi
\ifx \showarticletitle \undefined \def \showarticletitle #1{#1}   \fi
\ifx \showURL      \undefined \def \showURL       {\relax}        \fi
% invisible to TeX
\providecommand\bibfield[2]{#2}
\providecommand\bibinfo[2]{#2}
\providecommand\natexlab[1]{#1}
\providecommand\showeprint[2][]{arXiv:#2}

\bibitem[\protect\citeauthoryear{Bahdanau, Cho, and Bengio}{Bahdanau
  et~al\mbox{.}}{2014}]%
        {b27}
\bibfield{author}{\bibinfo{person}{Dzmitry Bahdanau},
  \bibinfo{person}{Kyunghyun Cho}, {and} \bibinfo{person}{Yoshua Bengio}.}
  \bibinfo{year}{2014}\natexlab{}.
\newblock \showarticletitle{Neural machine translation by jointly learning to
  align and translate}.
\newblock \bibinfo{journal}{\emph{arXiv preprint arXiv:1409.0473}}
  (\bibinfo{year}{2014}).
\newblock


\bibitem[\protect\citeauthoryear{Bui, Yu, and Jiang}{Bui et~al\mbox{.}}{2021}]%
        {b19}
\bibfield{author}{\bibinfo{person}{Nghi~DQ Bui}, \bibinfo{person}{Yijun Yu},
  {and} \bibinfo{person}{Lingxiao Jiang}.} \bibinfo{year}{2021}\natexlab{}.
\newblock \showarticletitle{Self-Supervised Contrastive Learning for Code
  Retrieval and Summarization via Semantic-Preserving Transformations}. In
  \bibinfo{booktitle}{\emph{Proceedings of the 44th International ACM SIGIR
  Conference on Research and Development in Information Retrieval}}.
  \bibinfo{pages}{511--521}.
\newblock


\bibitem[\protect\citeauthoryear{Gardner and Dorling}{Gardner and
  Dorling}{1998}]%
        {b33}
\bibfield{author}{\bibinfo{person}{Matt~W Gardner} {and} \bibinfo{person}{SR
  Dorling}.} \bibinfo{year}{1998}\natexlab{}.
\newblock \showarticletitle{Artificial neural networks (the multilayer
  perceptron)—a review of applications in the atmospheric sciences}.
\newblock \bibinfo{journal}{\emph{Atmospheric environment}}
  \bibinfo{volume}{32}, \bibinfo{number}{14-15} (\bibinfo{year}{1998}),
  \bibinfo{pages}{2627--2636}.
\newblock


\bibitem[\protect\citeauthoryear{Gu, Li, Gao, Wang, Zhang, Xu, and Lyu}{Gu
  et~al\mbox{.}}{2021}]%
        {b26}
\bibfield{author}{\bibinfo{person}{Wenchao Gu}, \bibinfo{person}{Zongjie Li},
  \bibinfo{person}{Cuiyun Gao}, \bibinfo{person}{Chaozheng Wang},
  \bibinfo{person}{Hongyu Zhang}, \bibinfo{person}{Zenglin Xu}, {and}
  \bibinfo{person}{Michael~R Lyu}.} \bibinfo{year}{2021}\natexlab{}.
\newblock \showarticletitle{CRaDLe: Deep code retrieval based on semantic
  Dependency Learning}.
\newblock \bibinfo{journal}{\emph{Neural Networks}}  \bibinfo{volume}{141}
  (\bibinfo{year}{2021}), \bibinfo{pages}{385--394}.
\newblock


\bibitem[\protect\citeauthoryear{Gu, Zhang, and Kim}{Gu et~al\mbox{.}}{2018}]%
        {b3}
\bibfield{author}{\bibinfo{person}{Xiaodong Gu}, \bibinfo{person}{Hongyu
  Zhang}, {and} \bibinfo{person}{Sunghun Kim}.}
  \bibinfo{year}{2018}\natexlab{}.
\newblock \showarticletitle{Deep code search}. In
  \bibinfo{booktitle}{\emph{2018 IEEE/ACM 40th International Conference on
  Software Engineering (ICSE)}}. IEEE, \bibinfo{pages}{933--944}.
\newblock


\bibitem[\protect\citeauthoryear{Guo, Fan, Ai, and Croft}{Guo
  et~al\mbox{.}}{2016}]%
        {b20}
\bibfield{author}{\bibinfo{person}{Jiafeng Guo}, \bibinfo{person}{Yixing Fan},
  \bibinfo{person}{Qingyao Ai}, {and} \bibinfo{person}{W~Bruce Croft}.}
  \bibinfo{year}{2016}\natexlab{}.
\newblock \showarticletitle{A deep relevance matching model for ad-hoc
  retrieval}. In \bibinfo{booktitle}{\emph{Proceedings of the 25th ACM
  international on conference on information and knowledge management}}.
  \bibinfo{pages}{55--64}.
\newblock


\bibitem[\protect\citeauthoryear{Haiduc, Bavota, Marcus, Oliveto, De~Lucia, and
  Menzies}{Haiduc et~al\mbox{.}}{2013}]%
        {b9}
\bibfield{author}{\bibinfo{person}{Sonia Haiduc}, \bibinfo{person}{Gabriele
  Bavota}, \bibinfo{person}{Andrian Marcus}, \bibinfo{person}{Rocco Oliveto},
  \bibinfo{person}{Andrea De~Lucia}, {and} \bibinfo{person}{Tim Menzies}.}
  \bibinfo{year}{2013}\natexlab{}.
\newblock \showarticletitle{Automatic query reformulations for text retrieval
  in software engineering}. In \bibinfo{booktitle}{\emph{2013 35th
  International Conference on Software Engineering (ICSE)}}. IEEE,
  \bibinfo{pages}{842--851}.
\newblock


\bibitem[\protect\citeauthoryear{Hinton, Srivastava, Krizhevsky, Sutskever, and
  Salakhutdinov}{Hinton et~al\mbox{.}}{2012}]%
        {b31}
\bibfield{author}{\bibinfo{person}{Geoffrey~E Hinton}, \bibinfo{person}{Nitish
  Srivastava}, \bibinfo{person}{Alex Krizhevsky}, \bibinfo{person}{Ilya
  Sutskever}, {and} \bibinfo{person}{Ruslan~R Salakhutdinov}.}
  \bibinfo{year}{2012}\natexlab{}.
\newblock \showarticletitle{Improving neural networks by preventing
  co-adaptation of feature detectors}.
\newblock \bibinfo{journal}{\emph{arXiv preprint arXiv:1207.0580}}
  (\bibinfo{year}{2012}).
\newblock


\bibitem[\protect\citeauthoryear{Hochreiter and Schmidhuber}{Hochreiter and
  Schmidhuber}{1997}]%
        {b6}
\bibfield{author}{\bibinfo{person}{Sepp Hochreiter} {and}
  \bibinfo{person}{J{\"u}rgen Schmidhuber}.} \bibinfo{year}{1997}\natexlab{}.
\newblock \showarticletitle{Long short-term memory}.
\newblock \bibinfo{journal}{\emph{Neural computation}} \bibinfo{volume}{9},
  \bibinfo{number}{8} (\bibinfo{year}{1997}), \bibinfo{pages}{1735--1780}.
\newblock


\bibitem[\protect\citeauthoryear{Huang, He, Gao, Deng, Acero, and Heck}{Huang
  et~al\mbox{.}}{2013}]%
        {b21}
\bibfield{author}{\bibinfo{person}{Po-Sen Huang}, \bibinfo{person}{Xiaodong
  He}, \bibinfo{person}{Jianfeng Gao}, \bibinfo{person}{Li Deng},
  \bibinfo{person}{Alex Acero}, {and} \bibinfo{person}{Larry Heck}.}
  \bibinfo{year}{2013}\natexlab{}.
\newblock \showarticletitle{Learning deep structured semantic models for web
  search using clickthrough data}. In \bibinfo{booktitle}{\emph{Proceedings of
  the 22nd ACM international conference on Information \& Knowledge
  Management}}. \bibinfo{pages}{2333--2338}.
\newblock


\bibitem[\protect\citeauthoryear{J{\"a}rvelin and
  Kek{\"a}l{\"a}inen}{J{\"a}rvelin and Kek{\"a}l{\"a}inen}{2002}]%
        {b32}
\bibfield{author}{\bibinfo{person}{Kalervo J{\"a}rvelin} {and}
  \bibinfo{person}{Jaana Kek{\"a}l{\"a}inen}.} \bibinfo{year}{2002}\natexlab{}.
\newblock \showarticletitle{Cumulated gain-based evaluation of IR techniques}.
\newblock \bibinfo{journal}{\emph{ACM Transactions on Information Systems
  (TOIS)}} \bibinfo{volume}{20}, \bibinfo{number}{4} (\bibinfo{year}{2002}),
  \bibinfo{pages}{422--446}.
\newblock


\bibitem[\protect\citeauthoryear{Kingma and Ba}{Kingma and Ba}{2014}]%
        {b10}
\bibfield{author}{\bibinfo{person}{Diederik~P Kingma} {and}
  \bibinfo{person}{Jimmy Ba}.} \bibinfo{year}{2014}\natexlab{}.
\newblock \showarticletitle{Adam: A method for stochastic optimization}.
\newblock \bibinfo{journal}{\emph{arXiv preprint arXiv:1412.6980}}
  (\bibinfo{year}{2014}).
\newblock


\bibitem[\protect\citeauthoryear{LeCun, Bottou, Bengio, and Haffner}{LeCun
  et~al\mbox{.}}{1998}]%
        {b7}
\bibfield{author}{\bibinfo{person}{Yann LeCun}, \bibinfo{person}{L{\'e}on
  Bottou}, \bibinfo{person}{Yoshua Bengio}, {and} \bibinfo{person}{Patrick
  Haffner}.} \bibinfo{year}{1998}\natexlab{}.
\newblock \showarticletitle{Gradient-based learning applied to document
  recognition}.
\newblock \bibinfo{journal}{\emph{Proc. IEEE}} \bibinfo{volume}{86},
  \bibinfo{number}{11} (\bibinfo{year}{1998}), \bibinfo{pages}{2278--2324}.
\newblock


\bibitem[\protect\citeauthoryear{Li, Wang, Wang, Yan, Xie, and Mei}{Li
  et~al\mbox{.}}{2016}]%
        {b11}
\bibfield{author}{\bibinfo{person}{Xuan Li}, \bibinfo{person}{Zerui Wang},
  \bibinfo{person}{Qianxiang Wang}, \bibinfo{person}{Shoumeng Yan},
  \bibinfo{person}{Tao Xie}, {and} \bibinfo{person}{Hong Mei}.}
  \bibinfo{year}{2016}\natexlab{}.
\newblock \showarticletitle{Relationship-aware code search for JavaScript
  frameworks}. In \bibinfo{booktitle}{\emph{Proceedings of the 2016 24th ACM
  SIGSOFT International Symposium on Foundations of Software Engineering}}.
  \bibinfo{pages}{690--701}.
\newblock


\bibitem[\protect\citeauthoryear{Li, Tarlow, Brockschmidt, and Zemel}{Li
  et~al\mbox{.}}{2015}]%
        {b30}
\bibfield{author}{\bibinfo{person}{Yujia Li}, \bibinfo{person}{Daniel Tarlow},
  \bibinfo{person}{Marc Brockschmidt}, {and} \bibinfo{person}{Richard Zemel}.}
  \bibinfo{year}{2015}\natexlab{}.
\newblock \showarticletitle{Gated graph sequence neural networks}.
\newblock \bibinfo{journal}{\emph{arXiv preprint arXiv:1511.05493}}
  (\bibinfo{year}{2015}).
\newblock


\bibitem[\protect\citeauthoryear{Ling, Wu, Wang, Pan, Ma, Xu, Liu, Wu, and
  Ji}{Ling et~al\mbox{.}}{2021}]%
        {b16}
\bibfield{author}{\bibinfo{person}{Xiang Ling}, \bibinfo{person}{Lingfei Wu},
  \bibinfo{person}{Saizhuo Wang}, \bibinfo{person}{Gaoning Pan},
  \bibinfo{person}{Tengfei Ma}, \bibinfo{person}{Fangli Xu},
  \bibinfo{person}{Alex~X Liu}, \bibinfo{person}{Chunming Wu}, {and}
  \bibinfo{person}{Shouling Ji}.} \bibinfo{year}{2021}\natexlab{}.
\newblock \showarticletitle{Deep Graph Matching and Searching for Semantic Code
  Retrieval}.
\newblock \bibinfo{journal}{\emph{ACM Transactions on Knowledge Discovery from
  Data (TKDD)}} \bibinfo{volume}{15}, \bibinfo{number}{5}
  (\bibinfo{year}{2021}), \bibinfo{pages}{1--21}.
\newblock


\bibitem[\protect\citeauthoryear{Linstead, Bajracharya, Ngo, Rigor, Lopes, and
  Baldi}{Linstead et~al\mbox{.}}{2009}]%
        {b12}
\bibfield{author}{\bibinfo{person}{Erik Linstead}, \bibinfo{person}{Sushil
  Bajracharya}, \bibinfo{person}{Trung Ngo}, \bibinfo{person}{Paul Rigor},
  \bibinfo{person}{Cristina Lopes}, {and} \bibinfo{person}{Pierre Baldi}.}
  \bibinfo{year}{2009}\natexlab{}.
\newblock \showarticletitle{Sourcerer: mining and searching internet-scale
  software repositories}.
\newblock \bibinfo{journal}{\emph{Data Mining and Knowledge Discovery}}
  \bibinfo{volume}{18}, \bibinfo{number}{2} (\bibinfo{year}{2009}),
  \bibinfo{pages}{300--336}.
\newblock


\bibitem[\protect\citeauthoryear{Lu, Sun, Wang, Lo, and Duan}{Lu
  et~al\mbox{.}}{2015}]%
        {b2}
\bibfield{author}{\bibinfo{person}{Meili Lu}, \bibinfo{person}{Xiaobing Sun},
  \bibinfo{person}{Shaowei Wang}, \bibinfo{person}{David Lo}, {and}
  \bibinfo{person}{Yucong Duan}.} \bibinfo{year}{2015}\natexlab{}.
\newblock \showarticletitle{Query expansion via wordnet for effective code
  search}. In \bibinfo{booktitle}{\emph{2015 IEEE 22nd International Conference
  on Software Analysis, Evolution, and Reengineering (SANER)}}. IEEE,
  \bibinfo{pages}{545--549}.
\newblock


\bibitem[\protect\citeauthoryear{Luong, Pham, and Manning}{Luong
  et~al\mbox{.}}{2015}]%
        {b28}
\bibfield{author}{\bibinfo{person}{Minh-Thang Luong}, \bibinfo{person}{Hieu
  Pham}, {and} \bibinfo{person}{Christopher~D Manning}.}
  \bibinfo{year}{2015}\natexlab{}.
\newblock \showarticletitle{Effective approaches to attention-based neural
  machine translation}.
\newblock \bibinfo{journal}{\emph{arXiv preprint arXiv:1508.04025}}
  (\bibinfo{year}{2015}).
\newblock


\bibitem[\protect\citeauthoryear{Lv, Zhang, Lou, Wang, Zhang, and Zhao}{Lv
  et~al\mbox{.}}{2015}]%
        {b1}
\bibfield{author}{\bibinfo{person}{Fei Lv}, \bibinfo{person}{Hongyu Zhang},
  \bibinfo{person}{Jian-guang Lou}, \bibinfo{person}{Shaowei Wang},
  \bibinfo{person}{Dongmei Zhang}, {and} \bibinfo{person}{Jianjun Zhao}.}
  \bibinfo{year}{2015}\natexlab{}.
\newblock \showarticletitle{Codehow: Effective code search based on api
  understanding and extended boolean model (e)}. In
  \bibinfo{booktitle}{\emph{2015 30th IEEE/ACM International Conference on
  Automated Software Engineering (ASE)}}. IEEE, \bibinfo{pages}{260--270}.
\newblock


\bibitem[\protect\citeauthoryear{McMillan, Grechanik, Poshyvanyk, Xie, and
  Fu}{McMillan et~al\mbox{.}}{2011}]%
        {b8}
\bibfield{author}{\bibinfo{person}{Collin McMillan}, \bibinfo{person}{Mark
  Grechanik}, \bibinfo{person}{Denys Poshyvanyk}, \bibinfo{person}{Qing Xie},
  {and} \bibinfo{person}{Chen Fu}.} \bibinfo{year}{2011}\natexlab{}.
\newblock \showarticletitle{Portfolio: finding relevant functions and their
  usage}. In \bibinfo{booktitle}{\emph{Proceedings of the 33rd International
  Conference on Software Engineering}}. \bibinfo{pages}{111--120}.
\newblock


\bibitem[\protect\citeauthoryear{Miller}{Miller}{1998}]%
        {b34}
\bibfield{author}{\bibinfo{person}{George~A Miller}.}
  \bibinfo{year}{1998}\natexlab{}.
\newblock \bibinfo{booktitle}{\emph{WordNet: An electronic lexical database}}.
\newblock \bibinfo{publisher}{MIT press}.
\newblock


\bibitem[\protect\citeauthoryear{Mitra, Diaz, and Craswell}{Mitra
  et~al\mbox{.}}{2017}]%
        {b24}
\bibfield{author}{\bibinfo{person}{Bhaskar Mitra}, \bibinfo{person}{Fernando
  Diaz}, {and} \bibinfo{person}{Nick Craswell}.}
  \bibinfo{year}{2017}\natexlab{}.
\newblock \showarticletitle{Learning to match using local and distributed
  representations of text for web search}. In
  \bibinfo{booktitle}{\emph{Proceedings of the 26th International Conference on
  World Wide Web}}. \bibinfo{pages}{1291--1299}.
\newblock


\bibitem[\protect\citeauthoryear{Shen, He, Gao, Deng, and Mesnil}{Shen
  et~al\mbox{.}}{2014}]%
        {b22}
\bibfield{author}{\bibinfo{person}{Yelong Shen}, \bibinfo{person}{Xiaodong He},
  \bibinfo{person}{Jianfeng Gao}, \bibinfo{person}{Li Deng}, {and}
  \bibinfo{person}{Gr{\'e}goire Mesnil}.} \bibinfo{year}{2014}\natexlab{}.
\newblock \showarticletitle{A latent semantic model with convolutional-pooling
  structure for information retrieval}. In
  \bibinfo{booktitle}{\emph{Proceedings of the 23rd ACM international
  conference on conference on information and knowledge management}}.
  \bibinfo{pages}{101--110}.
\newblock


\bibitem[\protect\citeauthoryear{Shuai, Xu, Liu, Yan, Xia, and Lei}{Shuai
  et~al\mbox{.}}{2020}]%
        {b4}
\bibfield{author}{\bibinfo{person}{Jianhang Shuai}, \bibinfo{person}{Ling Xu},
  \bibinfo{person}{Chao Liu}, \bibinfo{person}{Meng Yan}, \bibinfo{person}{Xin
  Xia}, {and} \bibinfo{person}{Yan Lei}.} \bibinfo{year}{2020}\natexlab{}.
\newblock \showarticletitle{Improving code search with co-attentive
  representation learning}. In \bibinfo{booktitle}{\emph{Proceedings of the
  28th International Conference on Program Comprehension}}.
  \bibinfo{pages}{196--207}.
\newblock


\bibitem[\protect\citeauthoryear{Sun, Liu, Yang, and Qian}{Sun
  et~al\mbox{.}}{2020}]%
        {b15}
\bibfield{author}{\bibinfo{person}{Zhensu Sun}, \bibinfo{person}{Yan Liu},
  \bibinfo{person}{Chen Yang}, {and} \bibinfo{person}{Yu Qian}.}
  \bibinfo{year}{2020}\natexlab{}.
\newblock \showarticletitle{PSCS: A Path-based Neural Model for Semantic Code
  Search}.
\newblock \bibinfo{journal}{\emph{arXiv preprint arXiv:2008.03042}}
  (\bibinfo{year}{2020}).
\newblock


\bibitem[\protect\citeauthoryear{Vaswani, Shazeer, Parmar, Uszkoreit, Jones,
  Gomez, Kaiser, and Polosukhin}{Vaswani et~al\mbox{.}}{2017}]%
        {b29}
\bibfield{author}{\bibinfo{person}{Ashish Vaswani}, \bibinfo{person}{Noam
  Shazeer}, \bibinfo{person}{Niki Parmar}, \bibinfo{person}{Jakob Uszkoreit},
  \bibinfo{person}{Llion Jones}, \bibinfo{person}{Aidan~N Gomez},
  \bibinfo{person}{{\L}ukasz Kaiser}, {and} \bibinfo{person}{Illia
  Polosukhin}.} \bibinfo{year}{2017}\natexlab{}.
\newblock \showarticletitle{Attention is all you need}. In
  \bibinfo{booktitle}{\emph{Advances in neural information processing
  systems}}. \bibinfo{pages}{5998--6008}.
\newblock


\bibitem[\protect\citeauthoryear{Wan, Shu, Sui, Xu, Zhao, Wu, and Yu}{Wan
  et~al\mbox{.}}{2019}]%
        {b14}
\bibfield{author}{\bibinfo{person}{Yao Wan}, \bibinfo{person}{Jingdong Shu},
  \bibinfo{person}{Yulei Sui}, \bibinfo{person}{Guandong Xu},
  \bibinfo{person}{Zhou Zhao}, \bibinfo{person}{Jian Wu}, {and}
  \bibinfo{person}{Philip~S Yu}.} \bibinfo{year}{2019}\natexlab{}.
\newblock \showarticletitle{Multi-modal attention network learning for semantic
  source code retrieval}.
\newblock \bibinfo{journal}{\emph{arXiv preprint arXiv:1909.13516}}
  (\bibinfo{year}{2019}).
\newblock


\bibitem[\protect\citeauthoryear{Xia, Bao, Lo, Kochhar, Hassan, and Xing}{Xia
  et~al\mbox{.}}{2017}]%
        {b25}
\bibfield{author}{\bibinfo{person}{Xin Xia}, \bibinfo{person}{Lingfeng Bao},
  \bibinfo{person}{David Lo}, \bibinfo{person}{Pavneet~Singh Kochhar},
  \bibinfo{person}{Ahmed~E Hassan}, {and} \bibinfo{person}{Zhenchang Xing}.}
  \bibinfo{year}{2017}\natexlab{}.
\newblock \showarticletitle{What do developers search for on the web?}
\newblock \bibinfo{journal}{\emph{Empirical Software Engineering}}
  \bibinfo{volume}{22}, \bibinfo{number}{6} (\bibinfo{year}{2017}),
  \bibinfo{pages}{3149--3185}.
\newblock


\bibitem[\protect\citeauthoryear{Xiong, Dai, Callan, Liu, and Power}{Xiong
  et~al\mbox{.}}{2017}]%
        {b23}
\bibfield{author}{\bibinfo{person}{Chenyan Xiong}, \bibinfo{person}{Zhuyun
  Dai}, \bibinfo{person}{Jamie Callan}, \bibinfo{person}{Zhiyuan Liu}, {and}
  \bibinfo{person}{Russell Power}.} \bibinfo{year}{2017}\natexlab{}.
\newblock \showarticletitle{End-to-end neural ad-hoc ranking with kernel
  pooling}. In \bibinfo{booktitle}{\emph{Proceedings of the 40th International
  ACM SIGIR conference on research and development in information retrieval}}.
  \bibinfo{pages}{55--64}.
\newblock


\bibitem[\protect\citeauthoryear{Xu, Yang, Liu, Shuai, Yan, Lei, and Xu}{Xu
  et~al\mbox{.}}{2021}]%
        {b13}
\bibfield{author}{\bibinfo{person}{Ling Xu}, \bibinfo{person}{Huanhuan Yang},
  \bibinfo{person}{Chao Liu}, \bibinfo{person}{Jianhang Shuai},
  \bibinfo{person}{Meng Yan}, \bibinfo{person}{Yan Lei}, {and}
  \bibinfo{person}{Zhou Xu}.} \bibinfo{year}{2021}\natexlab{}.
\newblock \showarticletitle{Two-Stage Attention-Based Model for Code Search
  with Textual and Structural Features}. In \bibinfo{booktitle}{\emph{2021 IEEE
  International Conference on Software Analysis, Evolution and Reengineering
  (SANER)}}. IEEE, \bibinfo{pages}{342--353}.
\newblock


\bibitem[\protect\citeauthoryear{Yao, Peddamail, and Sun}{Yao
  et~al\mbox{.}}{2019}]%
        {b17}
\bibfield{author}{\bibinfo{person}{Ziyu Yao},
  \bibinfo{person}{Jayavardhan~Reddy Peddamail}, {and} \bibinfo{person}{Huan
  Sun}.} \bibinfo{year}{2019}\natexlab{}.
\newblock \showarticletitle{Coacor: Code annotation for code retrieval with
  reinforcement learning}. In \bibinfo{booktitle}{\emph{The World Wide Web
  Conference}}. \bibinfo{pages}{2203--2214}.
\newblock


\bibitem[\protect\citeauthoryear{Ye, Xie, Zhang, Hu, Wang, and Zhang}{Ye
  et~al\mbox{.}}{2020}]%
        {b18}
\bibfield{author}{\bibinfo{person}{Wei Ye}, \bibinfo{person}{Rui Xie},
  \bibinfo{person}{Jinglei Zhang}, \bibinfo{person}{Tianxiang Hu},
  \bibinfo{person}{Xiaoyin Wang}, {and} \bibinfo{person}{Shikun Zhang}.}
  \bibinfo{year}{2020}\natexlab{}.
\newblock \showarticletitle{Leveraging code generation to improve code
  retrieval and summarization via dual learning}. In
  \bibinfo{booktitle}{\emph{Proceedings of The Web Conference 2020}}.
  \bibinfo{pages}{2309--2319}.
\newblock


\end{thebibliography}
\end{document}